\documentstyle[aps,12pt]{revtex}

\begin{document}
\title{Orbital moment in CoO and in NiO}
\author{R. J. Radwanski}
\address{Institute of Physics, Pedagogical University, 30-084 Krakow, Poland%
\\
\& Center for Solid State Physics, S$^{nt}$ Filip5,31-150Krakow,Poland.}
\author{Z. Ropka}
\address{Center for Solid State Physics, S$^{nt}$ Filip5,31-150Krakow,Poland.%
\\
email: sfradwan@cyf-kr.edu.pl, http://css-physics.edu.pl}
\maketitle

\begin{abstract}
The total, orbital and spin moment of the Co$^{2+}$ ion in CoO has been
calculated within the quasi-atomic approach with taking into account strong
correlations, crystal-field interactions and the intra-atomic spin-orbit
coupling. The orbital moment of 1.39 $\mu _{B}$ amounts at 0 K, in the
magnetically-ordered state, to more than 34\% of the total moment (4.01 $\mu
_{B}$) and yields the L/S ratio of 1.05, close to the experimental value.
The same calculations yield for NiO the orbital and total moment of 0.46 $%
\mu _{B}$ and 2.45 $\mu _{B}$, respectively.


\end{abstract}

{\bf 1. Introduction}

CoO and NiO attract a large attention of the magnetic community by more than
50 years. Despite of its simplicity (two atoms, $NaCl$ structure,
well-defined antiferromagnetism (AF) with T$_{N}$ of 290 K and 525 K,
respectively) and enormous theoretical and experimental works the consistent
description of its properties, reconciling its insulating state with the
unfilled 3$d$ band is still not reached \cite{1,2,3,4,5}.

The aim of this paper is to report the calculations of the magnetic moment
of CoO and NiO, its direction and the value as well as the spin and orbital
parts. In our approach, called Quantum Atomistic Solid-State theory
(QUASST), we attribute the moment of CoO/NiO to the Co$^{2+}$/Ni$^{2+}$
ions. We have calculated the moment of the Co$^{2+}$/Ni$^{2+}$ ion in the CoO%
$_{6}$/NiO$_{6}$ octahedral complex, its spin and orbital parts. The orbital
moment at 0 K as large as 1.38 $\mu _{B}$ and 0.46 $\mu _{B}$ has been found
in CoO and NiO. The approach used can be called the quasi-atomic approach %
\cite{6,7} as the starting point for the description of a solid is the
consideration of the atomic-like low-energy electronic structure of the
constituting atoms/ions, in the present case of the Co$^{2+}$/Ni$^{2+}$ ions.

{\bf 2. Theoretical outline}

We have treated the 7 outer electrons of the Co$^{2+}$ ion in Mott insulator
Co$^{2+}$O$^{2-}$ as forming the strongly-correlated electron system 3$d^{7}$%
. The correlations among electrons in the unfilled 3$d$ shell are taken by
two Hund's rules, that yield the ground-term quantum numbers $S$=3/2 and $L$%
=3 related to the ground term $^{4}F$ \cite{8,9}. Such the localized
highly-correlated electron system interacts in a solid with the charge and
spin surroundings. The charge surroundings has predominantly the octahedral
symmetry owing to the $NaCl$-type of structure of CoO and NiO\cite{4}.
Distortions are important for the detailed formation of the AF structure and
influence the spin and orbital moments but the most essential physical
interaction is the intra-atomic spin-orbit coupling. Our Hamiltonian for CoO
consists of two terms: the single-ion-like term $H_{d}$ of the 3$d^{7}$
system and the $d$-$d$ intersite spin-dependent term $H_{d-d}$, that is
indispensable for the formation of the magnetic state. The calculations
follow those, that we have performed for the description of FeBr$_{2}$ and
NiO \cite{6,7,10}. For the quasi-atomic single-ion-like Hamiltonian of the 3$%
d^{7}$ system we take into account the crystal-field interactions of the
octahedral symmetry (the octahedral CEF parameter $B_{4}$=-30 K), the
spin-orbit interactions with the spin-orbit coupling $\lambda _{s-o}$=-260 K
and the tetragonal distortion approximated by the term $B_{2}^{0}$= -5 K.
The single-ion states under the octahedral crystal field and the spin-orbit
coupling have been calculated by consideration of the Hamiltonian: 
\begin{equation}
H_{d}=B_{4}(O_{4}^{0}+5O_{4}^{4})+\lambda _{s-o}L\cdot S
\end{equation}%
The calculated single-ion states under the octahedral crystal field and the
spin-orbit coupling (the CoO$_{6}$ complex) are presented in Fig. 1(c). The
intersite spin-dependent interactions cause the (antiferro-)magnetic
ordering. They have been considered in the mean-field approximation with the
molecular-field coefficient {\it n} acting between magnetic moments $m_{d}$=(%
$L$+2$S$) $\mu _{B}$. Having the CEF\ interactions quantified, here $B_{4}$
mostly from optical studies, the value of $n$ in the Hamiltonian

\begin{equation}
H_{d-d}=n\left( -m_{d}\cdot m_{d}+\frac{1}{2}\left\langle
m_{d}^{2}\right\rangle \right)
\end{equation}

has been adjusted in order to reproduce the experimentally-observed Neel
temperature. The fitted value of $n$, for T$_{N}$ of 290 K, has been found
to be -2.43 meV/ $\mu _{B}^{2}$ (=42 T/$\mu _{B}$). It means that the Co-ion
moment in the magnetically-ordered state experiences the molecular field of
169 T (at 0 K).

As the result of calculations we get the low-energy electronic structure,
i.e. the energy of (2$L$+1)$\cdot $(2$S$+1) states and their eigenfunctions.
Having the energy of states we can calculate the free energy. It enables
calculations of thermodynamical properties. The ground-state eigenfunction
enables calculations of the total, orbital and spin moments.

In NiO the 8 outer electrons of the Ni$^{2+}$ ion, forming the
strongly-correlated electron system 3$d^{8}$, are described by $S$=1 and $L$%
=3 related to the ground term $^{3}F$ \cite{8}. For the calculations of the
quasi-atomic single-ion-like Hamiltonian of the 3$d^{8}$ system we take into
account the crystal-field interactions of the octahedral symmetry and the
spin-orbit coupling (the octahedral CEF parameter $B_{4}$=+2 meV and the
spin-orbit coupling constant $\lambda _{s-o}$=-41 meV \cite{8}, p. 399). The
preliminary results for NiO have been reported in \cite{10}.

{\bf 3. Results and discussion}

The calculated value of the total magnetic moment at 0 K in CoO and NiO in
the magnetically-ordered state amounts to 4.01 and 2.45 $\mu _{B}$,
respectively. In CoO it is built up from the spin moment $m_{s}$ of 2.62 $%
\mu _{B}$ ($S_{z}$=1.31) and the orbital moment $m_{o}$ of 1.39 $\mu _{B}$.
In NiO it is built up from the spin moment $m_{s}$ of 1.99 $\mu _{B}$ and
the orbital moment $m_{o}$ of 0.46 $\mu _{B}$.

The formation of the magnetic state and of the magnetic moment is associated
to the polarization of two conjugate states as is shown in Fig. 2. The
spin-like gap at 0 K amounts to 40 meV.

The orbital moment in CoO is quite substantial being more than 34\% of the
total moment. Our theoretical outcome, revealing the substantial orbital
moment is in nice agreement with the very recent experimental result of 3.98$%
\pm $0.06 $\mu _{B}$ for the Co moment \cite{3}. The magnetic x-ray
experiment has revealed the $L$/$S$ ratio of 0.95 \cite{11}- the calculated
by us values lead to $L$/$S$ ratio, in fact their $z$ components, of 1.05.
Also our theoretical result about the substantial orbital moment in NiO, is
in nice agreement with the very recent experimental result of 2.2$\pm $0.3 $%
\mu _{B}$ for the Ni moment at 300 K \cite{11,13}.

The effect of the small tetragonal distortion one can note comparing values
of the moment in the paramagnetic state shown in Fig. 2 and the moment of
the ground state in Fig. 1.

The calculated temperature dependence of the total moment, together with the
orbital and spin moments, is shown in Fig. 3. These moments disappear above T%
$_{N}$ - in the paramagnetic region the derived moment is the effective
moment, that bears the information about the operators $J^{2}$ or $S^{2}$.

We would like to point out that the evaluation of the orbital moment is
possible provided the spin-orbit coupling is taken into account. It confirms
the importance of the spin-orbit coupling for the description of the 3$d$%
-ion compounds. The magnetically-ordered state of CoO has lower energy than
the paramagnetic one by 380 J/mol (= 3.9 meV/ion) at 0 K. Of course, the
energies of magnetic and paramagnetic states become equal at $T_{N}$.

Finally, we would like to point out that our approach should not be
considered as the treatment of an isolated ion only - we consider the Co$%
^{2+}$ ion in the oxygen octahedron. The physical relevance of our
discussion to CoO is obvious - the $NaCl$ structure is built up from the
edge sharing Co$^{2+}$ octahedra and the same electronic structure occurs at
each Co site. The sign of the octupolar CEF parameter $B_{4}$ is directly
related to the oxygen anion surroundings and can be calculated from the
first principles. Its change of sign from the Co$^{2+}$ ion to Ni$^{2+}$ ion
is related to the sign reversal of the the octupolar moment of the charge
cloud of the 3$d^{7}$ and 3$d^{8}$ configurations, seen in Stevens factor $%
\beta $. The value of $B_{4}$ of -30 K in CoO is fully consistent with a
value of +260 K found in LaCoO$_{3}$ \cite{14} taking into account the
change of the valency and the average Co-O distance from 192 pm to 213 pm in
CoO.

{\bf 4. Conclusions}

The orbital and spin moment of the Co$^{2+}$ ion in CoO has been calculated
within the quasi-atomic approach taking into account strong correlations,
crystal-field interactions and the intra-atomic spin-orbit interactions by
the same approach as we have earlier used for NiO \cite{10}, for FeBr$_{2}$ %
\cite{7} and for LaCoO$_{3}$ \cite{14}. The orbital moment of 1.39 $\mu _{B}$
amounts at 0 K, in the magnetically-ordered state, to more than 34\% of the
total moment of 4.01 $\mu _{B}$. This value is much larger than 3 $\mu _{B}$
expected for 3 spin-holes \ in case of seven 3$d$ electrons in the Co$^{2+}$
ion. The presented approach explains in a very natural way the insulating
state of NiO - in fact, CoO and NiO are the best insulators. Our atomic-like
approach provides the discrete energy states for 3$d$ electrons in CoO and
calculates, apart of the spin moment, the orbital moment. Our studies
indicate that it is the highest time in solid-state physics to ''unquench''
the orbital moment in the description of 3$d$-atom containing compounds, the
more that it becomes visible in recent advanced experiments. The consistent
description of the low-energy electronic structure and the ordered magnetic
moment of four different compounds CoO, NiO, FeBr$_{2}$ and LaCoO$_{3}$
within the same atomistic approach we take as very strong argument for high
physical adequacy of our approach pointing out the existence of discrete
atomic-like states in 3$d$-atom containing compounds.


\begin{references}
\bibitem{1} V. J. Anisimov, J.\ Zaanen and O. K. Anderson, Phys. Rev. B {\bf %
44}, 943 (1991).

\bibitem{2} I. V Solovyev, A. I. Liechtenstein, and K. Terakura, Phys. Rev.
Lett. {\bf 80}, 5758 (1998).

\bibitem{3} T. Bredow and A. R. Gerson, Phys. Rev. B {\bf 61}, 5194 (2000).

\bibitem{4} W. Jauch, M. Reehuis, H. -J. Bleif, F. Kubanek, P. Pattison,
Phys. Rev. B {\bf 64}, 052102 (2001).

\bibitem{5} W. Jauch and M. Reehuis, Phys. Rev. B {\bf 65}, 125111 (2002).

\bibitem{6} R. J. Radwanski, R. Michalski, and Z. Ropka, Acta Phys. Pol. B 
{\bf 31}, 3079 (2000).

\bibitem{7} Z. Ropka, R. Michalski, and R. J. Radwanski, Phys. Rev. B {\bf 63%
}, 172404 (2001); http://xxx.lanl.gov/abs/cond-mat/0005502.

\bibitem{8} A. Abragam and B. Bleaney, in: {\it Electron Paramagnetic
Resonance of Transition Ions} (Clarendon, Oxford) 1970, ch.7.

\bibitem{9} C. J. Ballhausen, in: {\it Ligand Field Theory} (Mc-Graw-Hill
Comp.) 1962.

\bibitem{10} R. J. Radwanski and Z. Ropka, Acta Phys. Pol. A {\bf 97}, 963
(2000).

\bibitem{11} W. Neubeck, C. Vettier, F. de Bergevin, F. Yakhou, D. Mannix,
L. Ranno, and T. Chatterji, J.Phys.Chem. Solids {\bf 62}, 2173 (2001).

\bibitem{12} V. Fernandez, C. Vettier, F. de Bergevin, C.Giles, W. Neubeck,
Phys. Rev. B {\bf 57, } 7870 (1998).

\bibitem{13} W. Neubeck, C. Vettier, F. de Bergevin, F. Yakhou, D. Mannix,
O. Bengone, M. Alouani, and A. Barbier, Phys. Rev. B {\bf 63}, 134430 (2001).

\bibitem{14} Z. Ropka and R. J. Radwanski, Phys. Rev. B {\bf 67}, 172401
(2003); http://xxx.lanl.gov/abs/cond-mat/0301129.

F{\bf igure captions. \ Fig. 1.} The electronic structure of the Co$^{2+}$
ion in CoO. a) the free-ion $^{4}F$ term, b) the Co$^{2+}$ ion in the purely
octahedral crystal field, c) the electronic structure resulting from the
octahedral crystal field and the intra-atomic spin-orbit interactions.

Fig. 2. The splitting of the ground-state Kramers doublet, shown as the
lowest state in Fig. 1(c), in the magnetically-ordered state of CoO. Below T$%
_{N}$ a spin gap opens.

Fig. 3. Temperature dependence of the Co$^{2+}$-ion moment in CoO. At 0 K
the total moment of 4.01 $\mu _{B}$ is built up from the orbital and spin
moment of 1.39 and 2.62 $\mu _{B}$, respectively. The calculations have been
performed for the quasi-atomic parameters with the octahedral crystal-field
parameter B$_{4}$= -30 K, the spin-orbit coupling constant $\lambda _{s-o}$=
-260 K, the intersite spin-dependent interactions given by the
molecular-field coefficient $n$ = -42 T/ $\mu _{B}$ and the tetragonal
distortion parameter B$_{2}^{0}$ of -5 K.
\end{references}
\end{document}